\documentclass{osa-article}

\journal{osac}

\articletype{Research Article}

\usepackage{lineno}

\begin{document}

\title{Enhanced Polling and Infiltration for Highly-Efficient Electro-Optic Polymer-Based Mach-Zehnder Modulators}

\author{Iman Taghavi,\authormark{1,2} Razi Dehghannasiri,\authormark{2} Tianren Fan,\authormark{2} Alexander Tofini,\authormark{1} Hesam Moradinejad,\authormark{2} Ali. A. Efterkhar,\authormark{2} Sudip Shekhar,\authormark{1} Lukas Chrostowski,\authormark{1} Nicolas A. F. Jaeger,\authormark{1} , and Ali Adibi,\authormark{2}}

\address{\authormark{1} Electrical and Computer Engineering Department, Georgia Institute of Technology, 778 Atlantic Dr NW, Atlanta, GA 30332, USA\\
\authormark{2} Department of Electrical and Computer Engineering, University of British Columbia, 2332 Main Mall, Vancouver, B.C. V6T 1Z4, Canada}

\email{Corresponding author: ali.adibi@ece.gatech.edu} 



\begin{abstract} 
An ultra-narrow slot waveguide is fabricated for use in highly-efficient, electro-optic-polymer-based, integrated-optic modulators. Measurement results indicate that $V_\pi L$’s below 1.2 V.mm are possible for balanced Mach-Zehnder modulators using this ultra-narrow slot waveguide on a silicon-organic hybrid platform. Simulated $V_\pi L$’s of 0.35 V.mm have also been obtained. In addition to adapting standard recipes, we developed two novel fabrication processes for achieving miniaturized devices with high modulation sensitivity. To boost compactness and decrease the overall footprint, we use a fabrication approach based on air bridge interconnects on thick, thermally-reflowed, MaN 2410 E-beam resist protected by an alumina layer. To overcome the challenges of high currents and imperfect infiltration of polymers into ultra-narrow slots, we use a carefully designed, atomically-thin layer of TiO$_2$ as a carrier-barrier to enhance the polling efficiency of our electro-optic polymers. Additionally, finite-difference time-domain simulations are employed to optimize the effect of the thin layer of TiO$_2$. As compared to other, non-optimized, cases, our peak measured current is reduced by a factor of 3; scanning electron microscopy images also demonstrate that we achieve almost perfect infiltration. The anticipated increase in total capacitance due to the TiO$_2$ layer is shown to be negligible. In fact, applying our TiO$_2$ surface treatment to our ultra-narrow slot, allows us to obtain an improved phase shift efficiency ($\partial n / \partial V$) of $\sim$94\% for a 10 nm TiO$_2$ layer.
\end{abstract}

\section{Introduction}
Integrated photonic devices have received significant attention during the last decade thanks to their mass production, low power consumption, small footprints, and large optical bandwidths. To meet the demand for ultra-high-speed optical modulation, versatile structures have been studied for optical modulators as key building blocks in various photonic systems including microwave links, optical interconnects, and optical frequency comb generators. Advanced modulators, such as ones based on the plasma dispersion and electro-absorption effects, have been the subject of much recent research due to their promising characteristics. Nevertheless, they are not free of challenges since the maximum frequencies of both types of modulators are limited by their capacitances. \newline

Polymer-assisted modulators have emerged in various formats including single-mode strip waveguides, line-defect photonic crystal (PC) waveguides \cite{zhang2016high,zhang2013wide}, polymer-based waveguides \cite{enami2016analysis,enami2007hybrid}, slot waveguides\cite{koos2007nanophotonic}, athermal microrings \cite{qiu2016athermal}, and metal-polymer-silicon (Si)  waveguides \cite{sun2011design}. In all these structures, the goal is to maximize the field confinement to take full advantage of either phase shifts induced by Pockels-effect (i.e., $\chi^2$) polymers \cite{anderson2006high,chen2006subwavelength,qiu2016efficiently} or nonlinear Kerr-effect (i.e., $\chi^3$) polymers. Two widely used structures are Si-organic hybrid (SOH) and plasmonic-organic hybrid (POH) using which modulators with high bandwidths (> 100 GHz) \cite{chen1997demonstration, alloatti2014100} and ultra-low $V_\pi L$ \cite{palmer2014high,zhang2013polymer,witzens2010design,hochberg2007towards} have been fabricated so far. Possessing impressive characteristics including high switching speeds, low propagation losses, and low drive voltages \cite{palmer2014high} as compared with other modulators, EOP-based modulators (EOP: electro-optic polymer) offer a highly-resilient-to-radiation and CMOS-compatible alternative for next-generation on-chip optical signal processing systems \cite{taylor2005radiation}.\newline 

While plasmonic structures have better optical field confinement (i.e., optical field confinement and propagation happens at the metallic surface), SOH has lower loss and is more compatible with CMOS foundry processes. When making EOP-based modulators using conventional strip waveguides, made with Si or other materials, the EOP typically covers the waveguide, and only the evanescent field experiences the change in the refractive index due to the electro-optic effect of the polymer \cite{chen2011achieving}. Because of the large refractive index contrast between Si and EOP (n$_{\text{Si}}$ $\approx$
3.42 vs n$_{\text{EOP}}$ $\approx$  1.68), this type of SOH modulator suffers from poor optical-field/EOP overlap for conventional Si strip waveguides. On the other hand, in PC-based structures it is necessary for the EOP to penetrate the PC holes to modify the photonic bandgap properties. This necessitates a precise design and good fabrication robustness, which in turn imposes further costs and complexities. In polymer waveguides, interface layers, e.g., sol-gel silica layers, are commonly used to ensure that the DC electric field at the interface is uniform during polling. Each of the above-described devices has its shortcomings. In order to make low loss modulators with high modulation efficiencies, it has been proposed to use nano-slot Si waveguides in SOH structures. Vertical slots with slot widths from 75 nm to above 210 nm have been both theoretically and experimentally demonstrated to enhance the transverse electric (TE) mode intensity inside the Si nano-slots \cite{wang2011effective}. Indeed, in polymer-filled slots, the field overlap can be tailored to provide large electro-optic effects and, hence, allow one to make highly efficient modulators. \newline

Si slot-waveguide modulators utilizing EOP have not been free of challenges. Generally speaking, within limits, narrowing of a slot results in a better field overlap and hence, a better modulation efficiency. Slots narrower than 75 nm, however, have not been demonstrated yet due to fabrication difficulties. Even for the case of a 75 nm slot, the necessary high polling current degrades the polling process. The maximum tolerated polling field, before breakdown, is limited due to the high leakage current. Additionally, complete infiltration of the EOP inside the slot is, by itself, a difficult task. Both of these factors deteriorate the in-device, electro-optic coefficient (i.e. r$_{\text{33,in-device}}$), a crucial metric for the performance of a Mach-Zehnder modulator (MZM) and, hence, the overall modulation efficiency. These factors, i.e., polling and infiltration, have been widely investigated in Ref. \cite{szep2011poling,huang2012efficient,schulz2015mechanism}.\newline

To overcome these challenges, wider slots (up to 320 nm \cite{zhang2013wide}) have been employed in conjunction with PC-based waveguides. However, they suffer from design complexity and limited bandwidth. Recently, researchers have proposed the deposition of thin Alumina (Al$_2$O$_3$) as an interface for bulk layers of EOP on both Si and ITO (indium tin oxide) substrates \cite{schulz2015mechanism}. However, this study is limited to bulk epitaxially-grown layers of material, and the authors do not extend it to an applied modulator. Surface passivation necessity has also been studied elsewhere \cite{szep2011poling}, where the authors studied the effect of a titania (TiO$_2$) layer for passivation fabricated by ALD (atomic layer deposition). They concluded that addition of this layer not only decreases the leakage current but also increases the conductivity of Si pedestal and waveguides and hence improves polling efficiency. In this paper, we will focus on the other benefits of applying TiO$_2$ layers to very narrow slots.\newline

In this paper, we present our approach for a Si-organic hybrid platform that uses EOPs in ultra-narrow slot waveguides to achieve high modulation efficiencies. We report a series of surface preparation and infiltration techniques specifically for narrow vertical slots. We show that our approaches can improve the breakdown polling field and, hence, the r$_{\text{33,in-device}}$. We also demonstrate near-perfect void-free infiltration to decrease the polling current inside an ultra-narrow 40 nm-wide Si slot. Moreover, we present fabrication details for implementing air bridge interconnects to minimize the device footprints. While the focus in this paper is more on the platform development and not high-speed modulation, our approach does not impose any noticeable bandwidth limitations. In Section 2, we describe the design criteria to obtain various performance metrics as functions of structural factors. Fabrication details are extensively covered in Section 3 while Section 4 presents a discussion of the simulation and experimental results. The paper ends with conclusions in Section 5 drawn from the studies presented.

\section{Design Details}
Since the early demonstrations of SOH-based slotted and PC-waveguide modulators, these structures have always suffered from poor Si-organic interfaces causing incomplete infiltration and/or inefficient polling \cite{baehr2008nonlinear}. Ideally, we need to establish a strong DC electric field (i.e., a polling field) inside the polymer without drawing any DC current. Indeed, the leakage current flows between the sidewalls (i.e., between the cathode and the anode), preventing the complete polling of the chromophore molecules and may cause permanent damage/degradation to the polymer. While there is no standard recipe for surface functionalization, we have developed a customized surface treatment that allows us to maintain large electric fields inside the slots required for polling (100 V/cm), and at the same time to avoid reaching the breakdown field of the EOP. This method also helps polymer infiltration, i.e., it blocks air bubble formation at slot sidewalls, due to better adhesion to the sidewalls \cite{tang1997enhanced,enami2014enhanced,li2015poling,jouane2014unprecedented,sprave1996high,blum1998high}. \newline

\begin{figure}[h!]
\centering\includegraphics[width=1\linewidth]{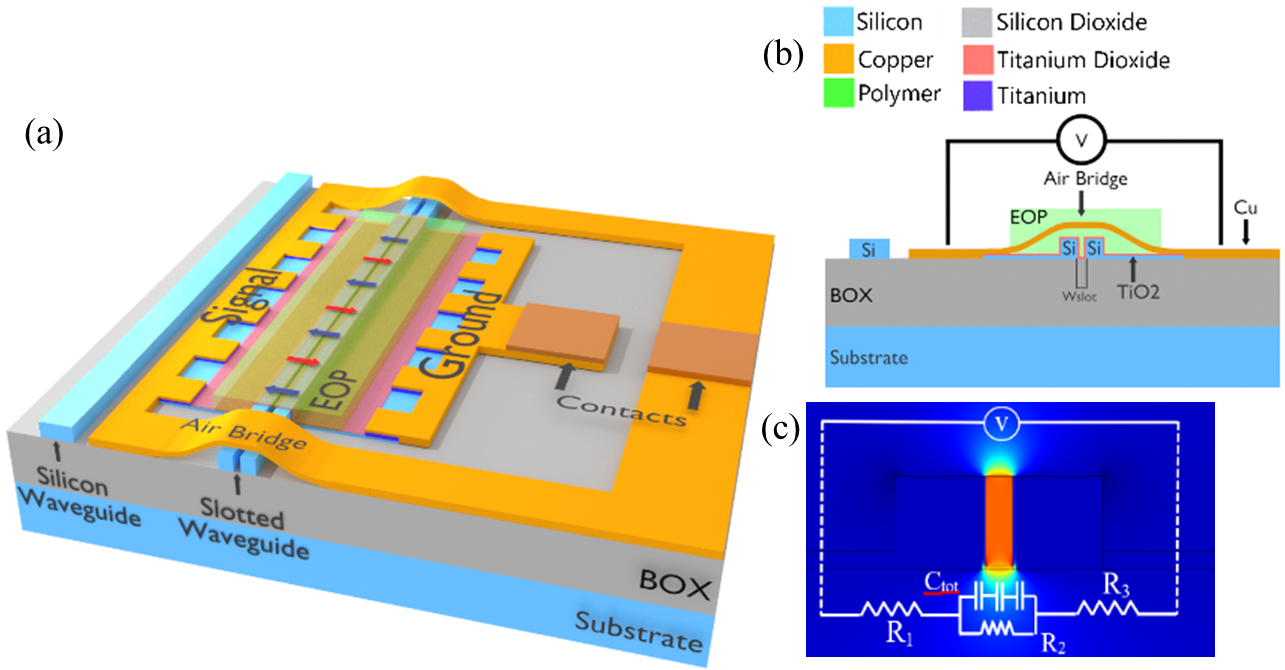}
\caption{(a) 3D Schematic of the MZM on SOH platform. (b) Cross-section of the device indicating the enhancing TiO$_2$. (c) Normalized, applied electric field (polling DC/RF) distribution. $R_1$, $R_3$ represent the resistances of the two pedestals, $R_2$ represents the combined resistance of the Si rails, EOP, and TiO$_2$ layers, and $C_{tot}$ represents the total capacitance of the Si rails, EOP, and TiO$_2$ layers.}
\label{Schematic}
\end{figure}

Our proposed SOH-based MZM is depicted in Fig. \ref{Schematic}. Among the many performance metrics for an MZM, the modulator sensitivity (i.e., the effective mode index change versus the applied voltage) is one of the most important properties, and it is given by 

\begin{equation}
S_p = \frac{\partial n_{\text{eff}}}{\partial V_{\text{in}}} = \frac{1}{2}n_e^3r_{\text{33,in-device}}\frac{\Gamma}{d_{\text{slot}}}
\label{Sp_eqn}
\end{equation}

where $n_e$ is the extraordinary refractive index of the EOP, $\Gamma$ is the field overlap integral, $d_{\text{slot}}$ is the slot width. The modulator drive voltage (V$_{\text{in}}$) and footprint are integrated into one single parameter called the modulation efficiency ($V_\pi L$), which is related to $S_{p}$ by $V_\pi L$ = $\lambda$/2$S_{p}$, where $\lambda$ is the optical wavelength, $V_\pi$ is the half-wave voltage (i.e., the applied voltage that results in an optical $\pi$ phase shift), and $L$ is the modulation length. While r$_{\text{33,in-device}}$ is directly related to the maximum tolerable polling field, $\Gamma$/$d_{\text{slot}}$ is the device geometry factor, which is obtained from both the optical (modal) and DC (polling) fields inside the slot. \newline

First, we need to evaluate the effect of our TiO$_2$ layer on the device speed. Being directly
related to the length, the capacitance of the thin TiO$_2$ and EOP layers are found to be CTiO$_2\sim$110 pF and CEOP$\sim$0.7 pF based on simple, parallel plate approximations within the slots. Here, we neglect the direct capacitance between two rails, i.e., the outer edge of the slot. The total equivalent slot capacitance is dominated by that of the EOP layer, since the contribution of the TiO$_2$ layer in series with the EOP layer is minimal. On the other hand, the thin TiO$_2$ layer does not contribute to the total resistance. As a result, the total RC is not increased drastically. Table \ref{ParameterTable} summarizes the modeling parameters and device geometries that we have used. As reported in \cite{enami2007hybrid}, adding TiO$_2$ may even reduce surface charges and, hence, reduce the pedestal resistance by functionalizing it; this would also help to improve the speed of the device. The added TiO$_2$ can slightly influence  the voltage drop across the slot estimated by

\begin{equation}
    V_{\text{EOP}} = V_a(\frac{W_{\text{EOP}}\cdot\rho_{\text{EOP}}}{2\cdot W_{\text{TiO$_2$}}\cdot \rho_{\text{TiO$_2$}} + W_{\text{EOP}}\cdot\rho_{\text{EOP}}}) \approx \theta V_a
\end{equation}

\noindent where $V_a$ is the total applied voltage across the device, $\theta$ is the structure factor, $\rho_{\text{EOP}}$ and $\rho_{\text{TiO$_2$}}$ are the polymer and TiO$_2$ resistivities, respectively, and the widths $W_{\text{EOP}}$ and $W_{\text{TiO$_2$}}$ are as shown in Fig.~\ref{Schematic}(b). Considering the proposed device in Fig.~\ref{Schematic}, we find $\theta$ $\sim$0.96. Evidently, one should not be concerned with the voltage drop across the thin oxide layer since it is only important when establishing a DC field for polling the EOP. The oxide interface forms an appropriate barrier for the carriers, which are responsible for the leakage current. 

\begin{table}[]
\centering
\resizebox{0.75\linewidth}{!}{
\tiny
\begin{tabular}{llllll}
\hline
\textbf{Parameter} & \textbf{Description} & \textbf{Value} & \textbf{Unit} & \textbf{Note} \\ \hline
$\theta$ & Structure factor & 0.96 & -  & For $d_{\text{slot}}$ = 40 nm; \\ 
 & & &  &    $W_{\text{TiO$_2$}}$ = 5 nm \\ \hline
$L$ & Arm length & 0.5 & mm &  \\ \hline
$r_{\text{33}}$ & EOP electrooptic coefficient & 100 & pm/v &  \\ \hline
$h$ & Waveguide height & 250 & nm & \\ \hline
$W_{\text{TiO$_2$}}$ & TiO$_2$ thickness & 5 & nm &  \\ \hline
$W_{\text{EOP}}$ & EOP thickness & 40/100 & nm &  \\ \hline
$W_{\text{Si}}$  & Silicon waveguide width& 250 & nm &  \\
  & (each rail) &  & & \\ \hline
$\mathcal{E}_{\text{TiO$_2$}}$ & TiO$_2$ relative permittivity & 86 &   &  \\ \hline
$\mathcal{E}_{\text{EOP}}$ & EOP permittivity &  2.8257 &  &   \\
&  &  (2.8257) & & &   \\ \hline
$\mathcal{E}_{\text{Si}}$ &  Si waveguide permittivity &  11.68  & &   \\ \hline
$\rho_{\text{TiO$_2$}}$ & TiO$_2$ resistivity & 1E5 & $\Omega m$ &  \\ \hline
$\rho_{\text{EOP}}$ & EOP resistivity & 1E6$\sim$7 & $\Omega m^*$ & \\ \hline
$n_e$ & EOP extraordinary  &  1.681 &  - &   @$\lambda = 1.55 \mu$m    \\ 
&refractive index &   &  &     \\ \hline
$n_0$ &  EOP ordinary  &  1.636 & - &  @$\lambda = 1.55 \mu$m  \\ 
 &  refractive index &  &  &  &   \\ \hline
$n_{\text{TiO$_2$}}$ &  TiO$_2$ refractive index & 2.61 & - & @$\lambda = 1.55 \mu$m  \\ \hline
$n_{\text{Si}}$ &  Si refractive index &  3.42 & -  &  @$\lambda = 1.55 \mu$m   \\ \hline
\end{tabular}
}
\caption{* EOP resistivity depends on the applied E field as well as the temperature. }
\label{ParameterTable}
\end{table}

We have simulated the effect of the TiO$_2$ thickness, among other factors, to see how it impacts various aspects of the device performance. Due to the electric field discontinuity at the slot interfaces, it is well-known that the TE electric field inside the slot (i.e., in the polymer infiltrated region) will be enhanced by $n_{\text{Si}}^2$/$n_{\text{EOP}}^2$ ($\approx$3.42$^2$/1.68$^2$ = 4.14 in our device) \cite{koos2007nanophotonic}. In the case of our thin TiO$_2$ layer, however, this ratio will be decreased to $n_{\text{Ti}O_2}^2$/$n_{\text{EOP}}^2$ ($\approx$ 2.61$^2$/1.68$^2$ = 2.41), i.e., by a factor of $n_{\text{Si}}^2$/$n_{\text{TiO}_2}^2$ $\approx$ 1.717. It is worth noting that the maximum polling efficiency is only possible when two conditions are met simultaneously: 1) optimal field overlap between the polling (DC) field and the optical TE (modal) field is obtained; 2) uniform void-free infiltration of the polymer inside the slot is achieved. The field overlap can be calculated by \cite{lou2013design,witzens2010design}  

\begin{equation}
    \Gamma = \int\frac{n_{\text{eff}}(\hat{E}_e)}{Z_0}|\hat{E}|^2dV \Big / \int \Re(\hat{E} \times \hat{H})dV
\end{equation}
where $n_{\text{eff}}$ is the effective mode index, $Z_0$ is the free space impedance, $V$ is the volume of integration, $\hat{E}$  and $\hat{H}$  are the modal electric and magnetic fields, respectively, $\hat{E}_e$  is the applied electric field (polling or RF) inside the slot, and $\Re$ represents the real-part operator.  Both integrals are calculated over the device volume $V$, while $n_{\text{eff}}(\hat{E}_e)$ is related to the polling field at any given point inside the slot by 

\begin{equation}
    n_{\text{eff}} = n_e - \frac{1}{2}n_e^3r_{33}\Gamma\hat{E}_e
\end{equation}

Iterative numerical methods are required to solve the coupled equations (3) and (4). Obviously, both $\hat{E}_x$ (the electric field component of the optical TE mode) and $\hat{E}_e$ play important roles in device performance; here, we first examine the role of $\hat{E}_x$. The effect of the DC polling field is inherent in $n_{\text{eff}}$, which is calculated along with $\hat{E}_x$ for every single point inside the slot using frequency domain simulations. Additionally, the denominator in Eq. (3) can be estimated by the total power transmitted in the medium.  \newline

\begin{figure}[h!]
\centering\includegraphics[width=1\linewidth]{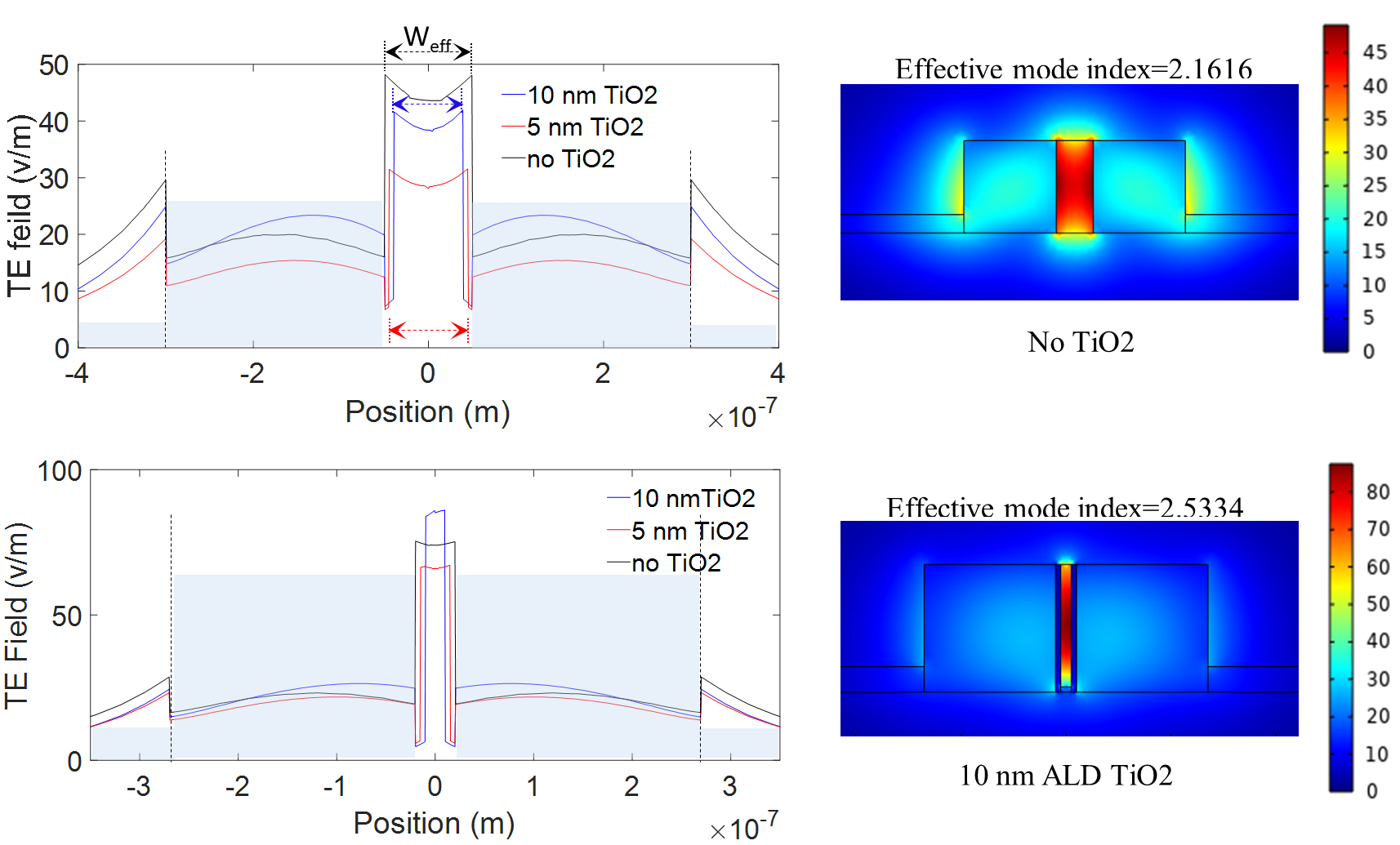}
\caption{Electric field norm of the slotted arm with and without a TiO$_2$ interface for: (top) 100 nm slot with no TiO$_2$ (bottom) 40 nm slot width. $W_{\text{eff}}$ is the effective slot width considering the space taken by TiO$_2$. }
\label{EField}
\end{figure}

As seen from Fig. \ref{EField} finite-difference eigenmode (FDE) simulation results can be used to model the effect of the TiO$_2$ layer on the modal field distribution. For the sake of comparison, we have repeated the calculations for a 100 nm-wide slot. The advantage of the 40 nm structures is suggested by the increased optical field intensity $|\hat{E}_x|$ in Fig. \ref{EField}, and, hence, the increased resulting modulation efficiency ($V_\pi L$). Figure \ref{EField} also shows that in case of the 100 nm-wide slot more of the field resides outside the waveguide, where it reaches a maximum of 30 V/$\mu$m. The maximum field inside the 100 nm-wide slot is 48 V/$\mu$m which is significantly lower than that in the 40 nm-wide slot. Since the TE field inside the 100 nm-wide slot is distributed less uniformly, the average TE field intensity is even lower. Also, adding TiO$_2$ smooths out the field, which is more pronounced in the 100 nm-wide slot since it starts out less-uniformly distributed, .\newline  

\begin{figure}[h!]
\centering\includegraphics[width=1\linewidth]{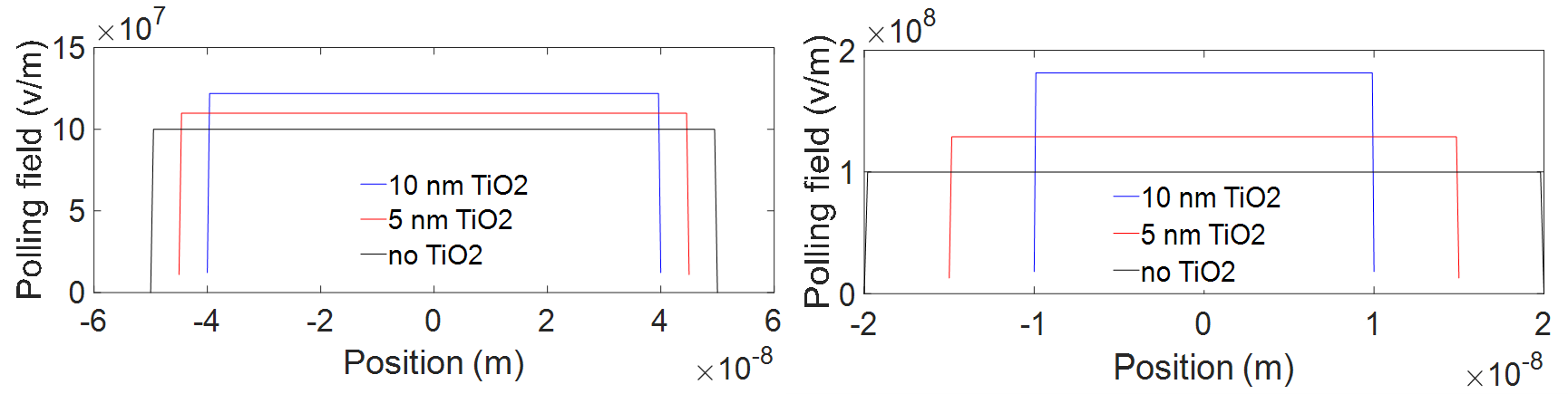}
\caption{Normalized (to 1 V applied) electric field distribution inside slots of width (left) 100 nm and (right) 40 nm. Higher polling field along with better field tolerance due to carrier-barrier behavior of the oxide interface result in higher $r_{\text{33,in-device}}$ and, hence, better modulation efficiency.}
\label{EField_Zoom}
\end{figure}

Fig. \ref{EField_Zoom}, shows the simulation results for $\hat{E}_e$, calculated halfway up the rails (i.e., 110 nm above the BOX layer Fig.~\ref{Schematic}(b)). Together with Fig. \ref{EField}, one may identify two immediate effects of the added TiO$_2$ layer  as: 1) the polling field is enhanced by factors of 30\% and 80\% for 5 nm and 10 nm of TiO$_2$ on each side of the 40 nm slot, respectively (that is important because the SEO125B polymer requires a minimum polling field of 100 V/$\mu$m for effective chromophore alignment) and 2) the uniformity of the modal field throughout the slot could be improved as shown in  Fig. 2. This effect was previously reported in Ref. [23], where the authors employed an interface layer of TiO$_2$ to enhance polling-field uniformity albeit for non-slots electro-optic modulators.\newline

\begin{figure}[h!]
\centering\includegraphics[width=1\linewidth]{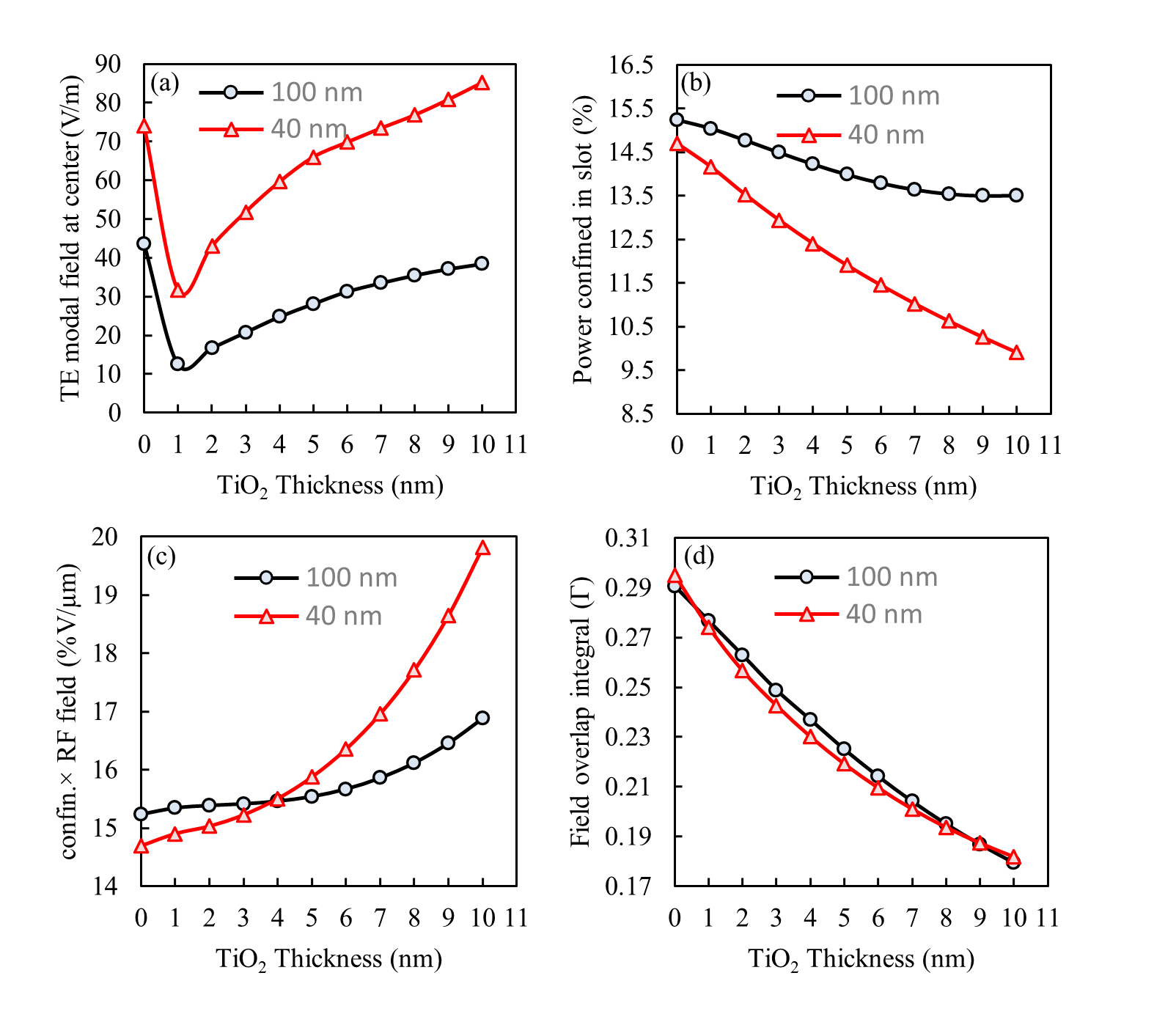}
\caption{Electrical (polling/RF) and optical field characteristics for slots with 100 nm and 40 nm widths: (a) Modal field intensity at the slot center; (b) Power confinement ratio inside the slot relative to the entire propagation medium; (c) Power confinement ratio multiplied by the electric field, $\kappa |\hat{E}_{ave}|$; and (d) Field overlap integral, $\Gamma$.}
\label{Gamma}
\end{figure}

Figure \ref{Gamma}(a) shows the modal field intensity measured at the midpoint of the slot, where the infiltration of the EOP is most efficient. The thicker the buffer layer, the higher the TE field intensity. Figure \ref{Gamma}(b) illustrates the power confinement factor ($\kappa$) in the slot. Depositing more TiO$_2$ into a smaller slot will result in less available space for the EOP. Figures \ref{Gamma}(c) and \ref{Gamma}(d) show a more intuitive figure of merit, $\kappa |\hat{E}_{\text{avg}}|$, where $|\hat{E}_{\text{avg}}|$ is the average value of $|\hat{E}|$ within the slot. The thicker the TiO$_2$ used, the higher the value of $\kappa |\hat{E}_{\text{avg}}|$. Yet, as shown in Fig. \ref{Gamma}(d), $\Gamma$ is reduced by adding more oxide. It is worth noting that, while both are optical-electrical field overlap factors, $\Gamma$ is different than $\kappa |\hat{E}_{\text{avg}}|$, and it is the parameter that eventually defines the performance of the device.

\section{Fabrication Methods}
\subsection{Device}
\begin{figure}[h!]
\centering\includegraphics[width=1\linewidth]{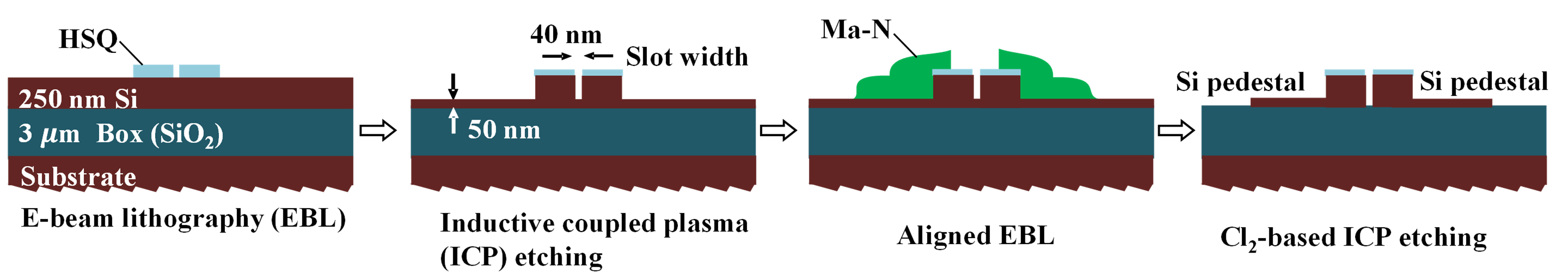}
\caption{Fabrication flow diagram of the MZM with air bridge interconnect and surface treatment steps}
\label{FabFlow}
\end{figure}

While 120 nm slots are commonly used in slotted-waveguide SOH modulators, fabricating narrower slots (e.g., 40 nm slots) is much more challenging. Next, we describe those challenges and provide our detailed process flow steps. The modulator fabrication flow diagram is shown in Fig. \ref{FabFlow}. We begin with a 250 nm / 3 $\mu$m BOX SOI chip. The slotted waveguide was formed by E-beam lithography (EBL) and dry etching. A JEOL JBX-9300FS EBL System, with a resolution of 20 nm at 100 pA beam current, was utilized to create ultra-narrow slots. For the E-beam resist, our silicon slot was patterned using HSQ 6\%, whereas MaN 2403 was used for our pedestals. It is important to use a gentle, step-by-step etching process (etch rate equal to $\sim$75 nm/min) to make sure that vertical sidewalls are obtained with minimal surface roughness, especially when making such small slots that will have low optical losses.  To succeed in this crucial step, a chlorine-based etch process was used in an inductively coupled plasma (ICP) etching system. \newline

The pedestals are designed to have low ohmic losses and to carry both polling and modulation signals to the slot sidewalls. After the formation of the MZM, 150 $\mu$m x 150 $\mu$m contact pads of are fabricated in a third EBL step, with ZEP as the E-beam resist. A Denton Explorer E-beam evaporation system is employed for metallization in which a titanium (Ti) layer (10 nm) followed by a copper (Cu) layer (250 nm) are deposited under moderate vacuum conditions (pressure < $3 \times 10^{-6}$ Torr). The metal lift-off is processed using 1165 at 80 °C overnight for gentle, crack-free interconnects. Then the polymer is prepared and spin-coated according to our recipe (discussed later). The EOP on top of the contacts, formed in the last step, is removed at an elevated temperature (> 130 °C) prior to polling. Additionally, we use tunable directional couplers instead of more common multi-mode interferometer (MMI) couplers or Y-splitters, because they allow for balancing of the MZM to maximize the extinction ratio. The 100 nm gap of the directional couplers is protected by an additional layer of flowable oxide (FoX), which otherwise could be filled with polymer during the EOP-deposition process.\newline

\begin{figure}[h!]
\centering\includegraphics[width=1\linewidth]{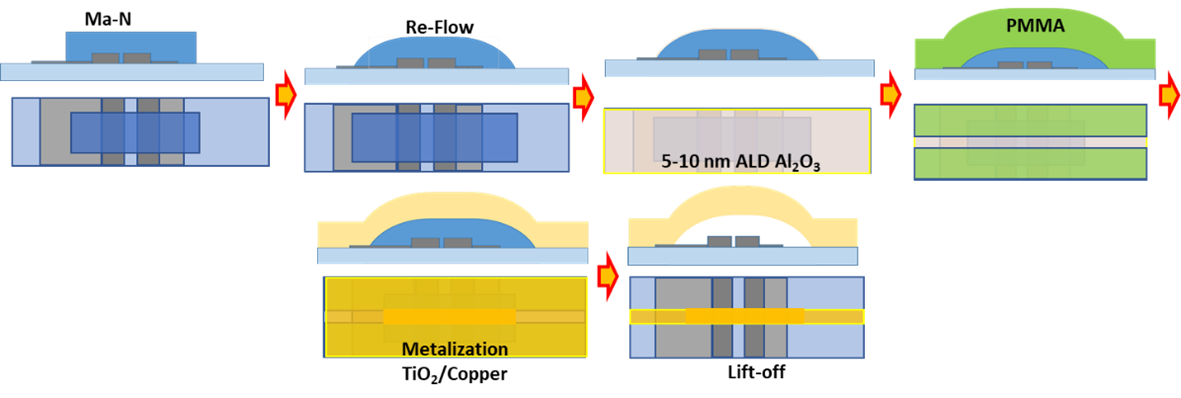}
\caption{Fabrication steps for the air-bridge interconnect which includes two EBL, re-flow, and metallization steps and one ALD step (both cross-section and top views are shown for each step). PMMA and Ma-N 2410 E-beam resists is used as the metallization mask and the bridge support, respectively.}
\label{Bridge}
\end{figure}

\begin{figure}[h!]
\centering\includegraphics[width=0.5\linewidth]{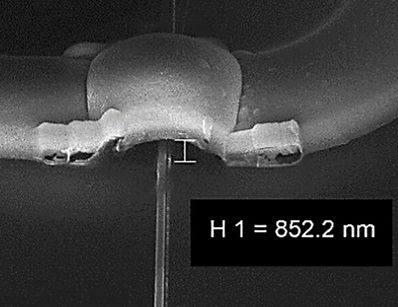}
\caption{SEM image of the fabricated air bridge after metallization and the removal of the MaN 2410. }
\label{SEM}
\end{figure}

To enable contacts for applying electrical signals (both RF and polling), vertical vias made of conducting materials, e.g., tungsten, are usually employed on top of the samples \cite{korn2014electro} and insulated with SiO$_2$. Indeed, achieving metallic contact with the inner parts of the MZM is a challenge due to geometry limitations and demand for the optimum device footprints. Compared to conventional vias, we found that polymer infiltration could be more efficient using our air-bridge-over-photonic-waveguide interconnect approach shown in Fig. \ref{Bridge}. This process starts by spin-coating a 1.3 $\mu$m-thick E-beam resist (MaN 2410) on top of the already formed MZM. One should minimize electrical resistance of the air bridge, which directly affects the polling efficiency. This critical characteristic is achieved by a reflow process in which the sharp corners of the processed rectangles are reduced from $\sim$ 90° to $\sim$ 70°. We have determined that reflowing for 2 minutes at $\sim$ 145 °C generates the best outcome. Employing the FDTD simulations, we have determined the geometry requirements of the air-bridge for minimal optical loss. Accordingly, our simulations show that minimum distances of 450 nm from the top of waveguide/slot region and 600 nm from each side of the waveguide are necessary to avoid considerable optical loss caused by the air bridge. In the next step, atomically thin layer of alumina is coated using ALD followed by one further step of E-beam lithography, copper sputtering and subsequent lift-off using Polymethyl methacrylate (PMMA) for metallization. Despite its nanoscale size, our developed process flow resulted in both mechanically-stable and reproducible results for the air-bridge. Depicted in Fig. \ref{SEM} is the scanning electron microscopy (SEM) image of the fabricated metal air bridge on top of the waveguide.

\subsection{Material}
The most challenging step in making narrow-slot, EOP-filled waveguides is achieving void-free EOP infiltration of the slots \cite{szep2011poling,huang2012efficient,schulz2015mechanism}. In addition to the benefits mentioned previously, our TiO$_2$ surface treatment can help in this regard. We have tested various mechanisms to enhance the polymer-Si interface. First, we ran a customized oxygen plasma cleaning step on the surface (using advanced vacuum reactive ion etching (RIE)). Excessive plasma cleaning, however, damages the waveguides and should be avoided. As illustrated in Fig. 8, we deposited a thin layer (5 nm) of TiO$_2$ everywhere, including inside the slot, using a Cambridge atomic layer deposition (ALD) system. ALD is known to be the most precise deposition method for thin oxide films with desired thicknesses. ALD also produces a high-quality interface between the Si slot sidewalls and the oxide (TiO$_2$), as compared to alternative methods such as plasma-enhanced chemical vapor deposition (PECVD), and is known to provide conformal coatings even in confined spaces such as under our air-bridges. The ALD process temperature was set to 150 °C and a layer of $\approx$ 0.7 Å was deposited in each cycle. The pulse and purge times in one ALD cycle were carefully monitored to maintain the quality of the oxide coating.\newline

\begin{figure}[h!]
\centering\includegraphics[width=1\linewidth]{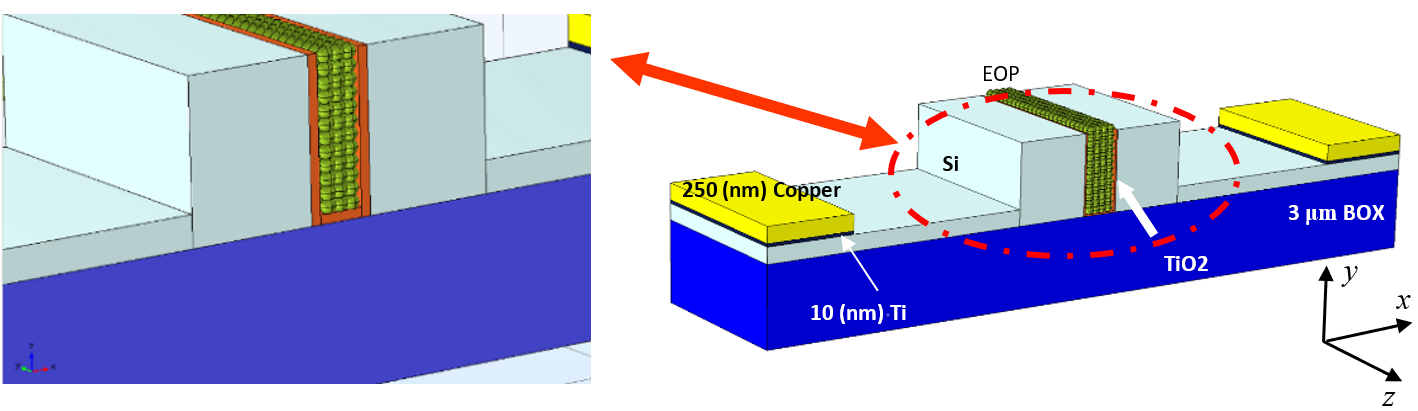}
\caption{Perspective view of a schematic of the slotted waveguide cross-section. Shown in red is the deposited, 5 nm TiO$_2$ layer (a zoomed view is shown on the left).}
\label{SlotWaveguide}
\end{figure}

As compared to the standard recipe for SEO125B (6.72\% by weight according to Soluxra Inc.), we deliberately diluted the solution by increasing the volume ratio of the solvent (Dibromomethane) to SEO125B by a factor of $\sim$ 1.5. We have also devised a method called “rotating bubble” for spin coating the EOP as a replacement for the conventional “dropping” method prescribed by the manufacturer. In this method, after dispensing a small amount of EOP onto an unused portion of the chip using a bulb pipette, by continuing to apply pressure to the dispensing pipette’s bulb a small bubble of 2$\sim$3 mm diameter is formed at the end of the pipette. This bubble can be used to deposit a thin layer of the polymer (a few hundred nanometers thick) across the surface of the chip by “pulling” the bubble through moving the pipette. Simply dropping the EOP onto the chip’s surface leaves a bulky polymer that not only causes an additional loss but also complicates the opening of the windows and causes the EOP to develop harmful cracks. Using our optimized process, we have successfully developed a recipe for the spin coating that decreases the thickness of the EOP on top of the sample from > 3 $\mu$m to $\sim$450 nm. During the spin-coating process, due to the viscosity of the EOP, the initial drop does not spread significantly but the thin layer, formed by our technique, flows enough to increase the uniformity of the EOP across the critical portions of the chip. The thickness obtained is adequate since the waveguide height is only 250 nm. It should be noted that our techniques for EOP dilution and “rotating deposition” also help to decrease the additional loss imposed on the input/output grating couplers by the polymer. Consequently, as will be demonstrated later, this should result in better extinction ratios (ERs), which are determined by the optical loss asymmetry of the two arms. Another benefit of our EOP deposition approach is the infiltration inside the slot. By waiting a few minutes after spinning, we allow for the polymer to be pulled into the by the capillary effect. Otherwise, the polymer may not penetrate the slot uniformly. Employing these techniques, we have succeeded in obtaining better polymer infiltrations compared to all reported alternatives \cite{szep2011poling,huang2012efficient,schulz2015mechanism}. \newline

\begin{figure}[h!]
\centering\includegraphics[width=1\linewidth]{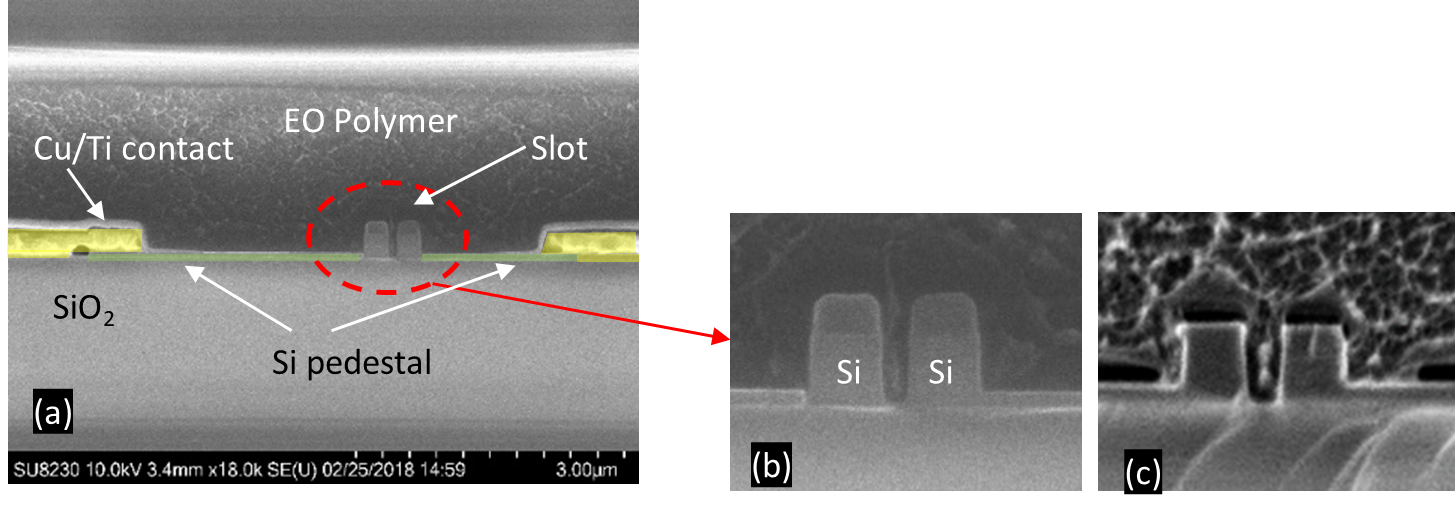}
\caption{(a) A false-colored SEM of a 40 nm slotted waveguide, with 5 nm TiO$_2$, infiltrated with SEO125B. (b) A magnified image of the slot shown in (a) demonstrates the excellent infiltration and surface adhesion. (c) false-colored SEM of a 40 nm slotted waveguide with an ALD-deposited, 5 nm thick, Al$_2$O$_3$ layer, as was done in \cite{li2015poling} }
\label{SEM_Cross}
\end{figure}

The coated sample is then heated overnight in a vacuum oven at $\sim$75 °C to remove leftover air gaps/voids from the slot. Figure 9 shows SEM images of the slot with (Figs. \ref{SEM_Cross}(a) and \ref{SEM_Cross}(b)) and without (Fig. \ref{SEM_Cross}(c)) enhanced polymer infiltration method. Schulz et al. have previously studied the effect of an additive layer of alumina (Al$_2$O$_3$) on the non-slotted waveguide \cite{schulz2015mechanism}. For the sake of comparison, we have also deposited 5 nm of Al$_2$O$_3$ and monitored the infiltration/adhesion of EOP into the slot/Si sidewalls. As seen in Fig. \ref{SEM_Cross} (a), almost-perfect infiltration of the EOP is achieved for the TiO$_2$ deposited sample whereas for the Al$_2$O$_3$ case (Fig. \ref{SEM_Cross} (c)) only a portion of the slot volume is infiltrated. Figure \ref{SEM_Cross} demonstrates the improvement obtained for the interface between TiO$_2$ and SEO125B as compared to that between conventional Al$_2$O$_3$ and SEO125B.

\section{Results and Discussion}
\begin{figure}[h!]
\centering\includegraphics[width=1\linewidth]{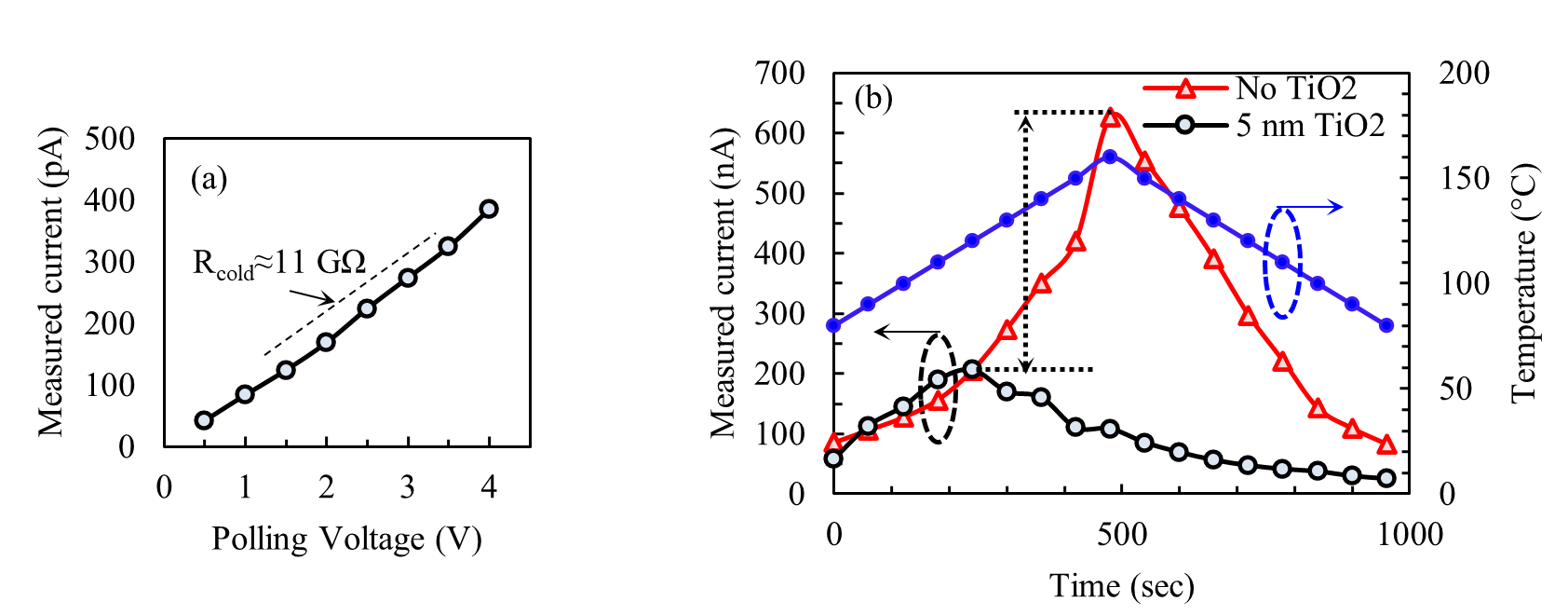}
\caption{(a) Cold (i.e., room temperature) resistance measurements and (b) the measured current versus time for the temperature ramp used during the polling procedure. Less than one-third of the peak measured current is observed when the thin-TiO$_2$ protection layer is applied. }
\label{Polling}
\end{figure}

The linear electro-optic (EO) effect (the Pockels effect) can be induced in the EOP by orienting its chromophore molecules in such a manner that the optical center of symmetry of the polymer is removed \cite{lee2011recent}. Generally speaking, the larger the polling field inside the slot, the larger the induced EO \cite{lee2011recent} effect. However, the maximum polling field is limited by the polymer breakdown. SEO125B has a standard polling recipe in which a polling field of at least 100 V/$\mu$m should be applied to the polymer. To obtain this field, we minimize the resistance of the 50 nm pedestal and the waveguide rail compared to the thin film EOP. Then, a temperature ramp of 10 °C/min is applied, while maintaining the field, to reach a state in which chromophores can be readily aligned. The temperature is increased from 80 °C to 150 °C, given that the glass transition temperature of the polymer, Tg, is $\sim$ 145 °C. The applied current drawn is recorded throughout this process. Figure \ref{Polling}(a) depicts the cold resistance while Fig. \ref{Polling}(b) shows the measured current versus temperature, for the temperature ramp used during the polling procedure, for a 40 (nm) slot waveguide with and without the TiO$_2$ layer. \newline

\begin{figure}[h!]
\centering\includegraphics[width=0.75\linewidth]{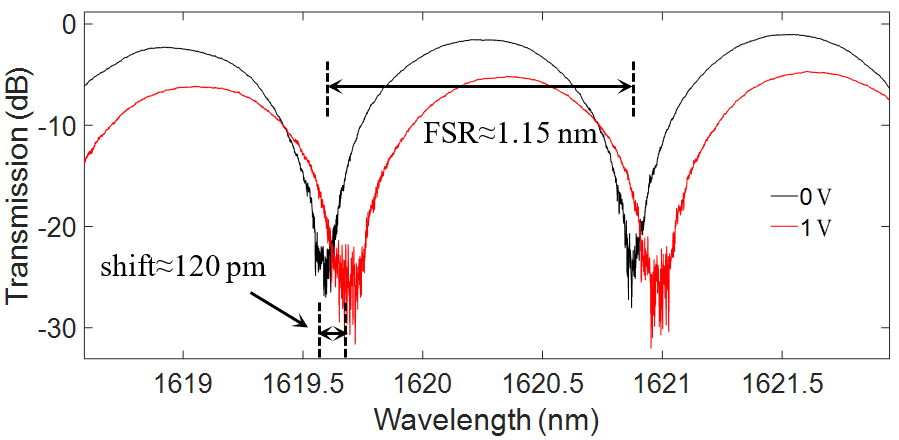}
\caption{Transmission spectra of a slotted, SOH MZM with a TiO$_2$ passivated surface, a 40 nm slot width, and a 1 mm arm length. }
\label{Transmission}
\end{figure}

Evidently, our TiO$_2$ surface functionalization results in a peak current, which is less than one-third of that in the non-treated surface case.  Considering a surface area of $250\times10^{-9}\times3\times10^{-3}$ m$^2$, the peak current of 206 nA observed for the TiO$_2$-protected case (Fig. \ref{Polling}(b)) is equivalent to a current density of $\approx$ 275 A/m$^2$. This number is within the range that was reported in Ref. \cite{wang2011effective}, albeit with a 75 nm slot size. Since our device has a 40 nm slot, it is anticipated that a higher $S_p$, and hence modulation gain, should be achieved. We believe that this improvement is due to the thin TiO$_2$ layer acting as a “carrier-stopper”, along with our improved EOP infiltration. In other words, the high r33,in-device reported in \cite{wang2011effective} (associated with the higher optical field overlap thanks to PC-based structure) could be even more enhanced using our proposed TiO$_2$ addition layer.\newline

\begin{figure}[h!]
\centering\includegraphics[width=1\linewidth]{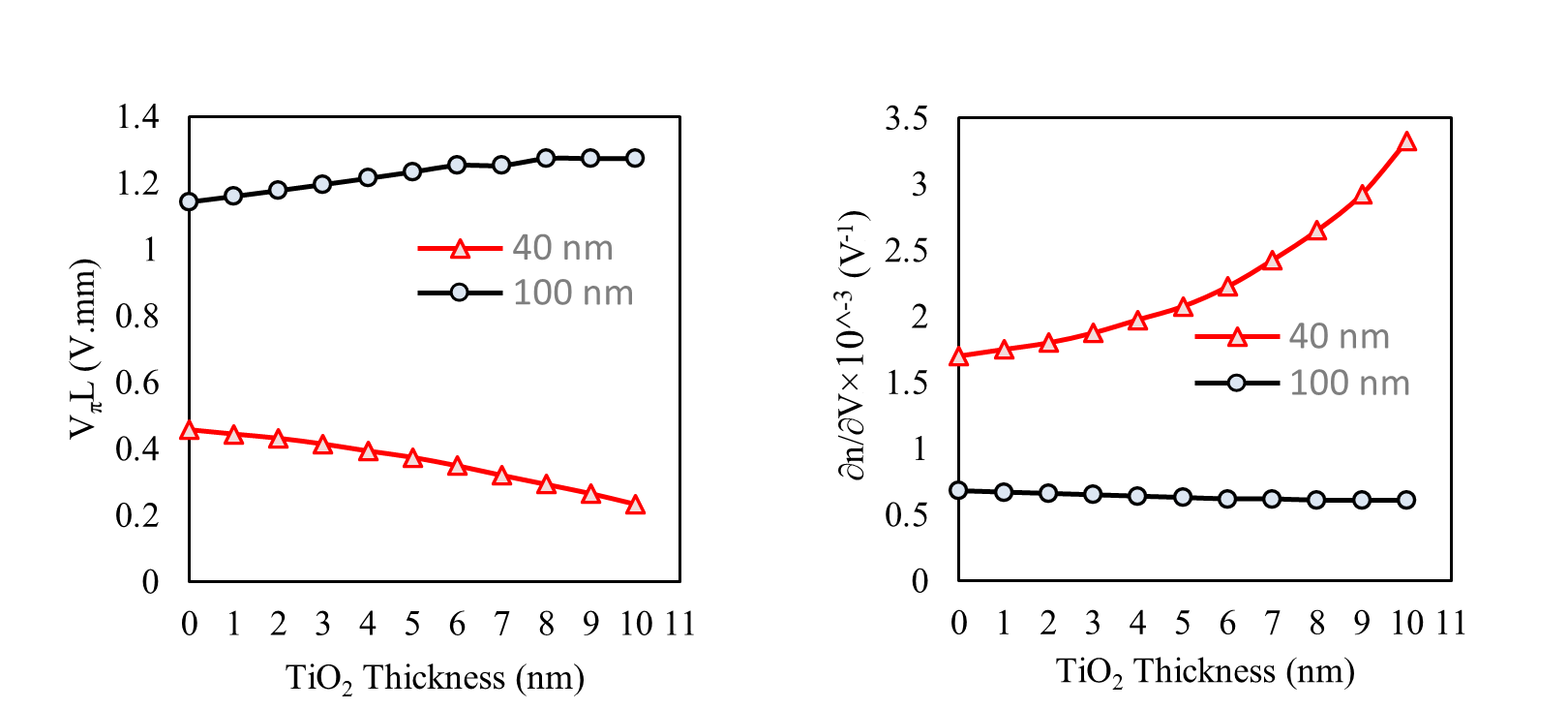}
\caption{(a) Effective mode index changes versus the applied voltage ($S_p$) and (b) modulation efficiency ($V_\pi L$) as functions of TiO$_2$ thickness for devices with 100 nm and 40 nm slot widths. }
\label{ModeVsThickness}
\end{figure}

\begin{figure}[h!]
\centering\includegraphics[width=1\linewidth]{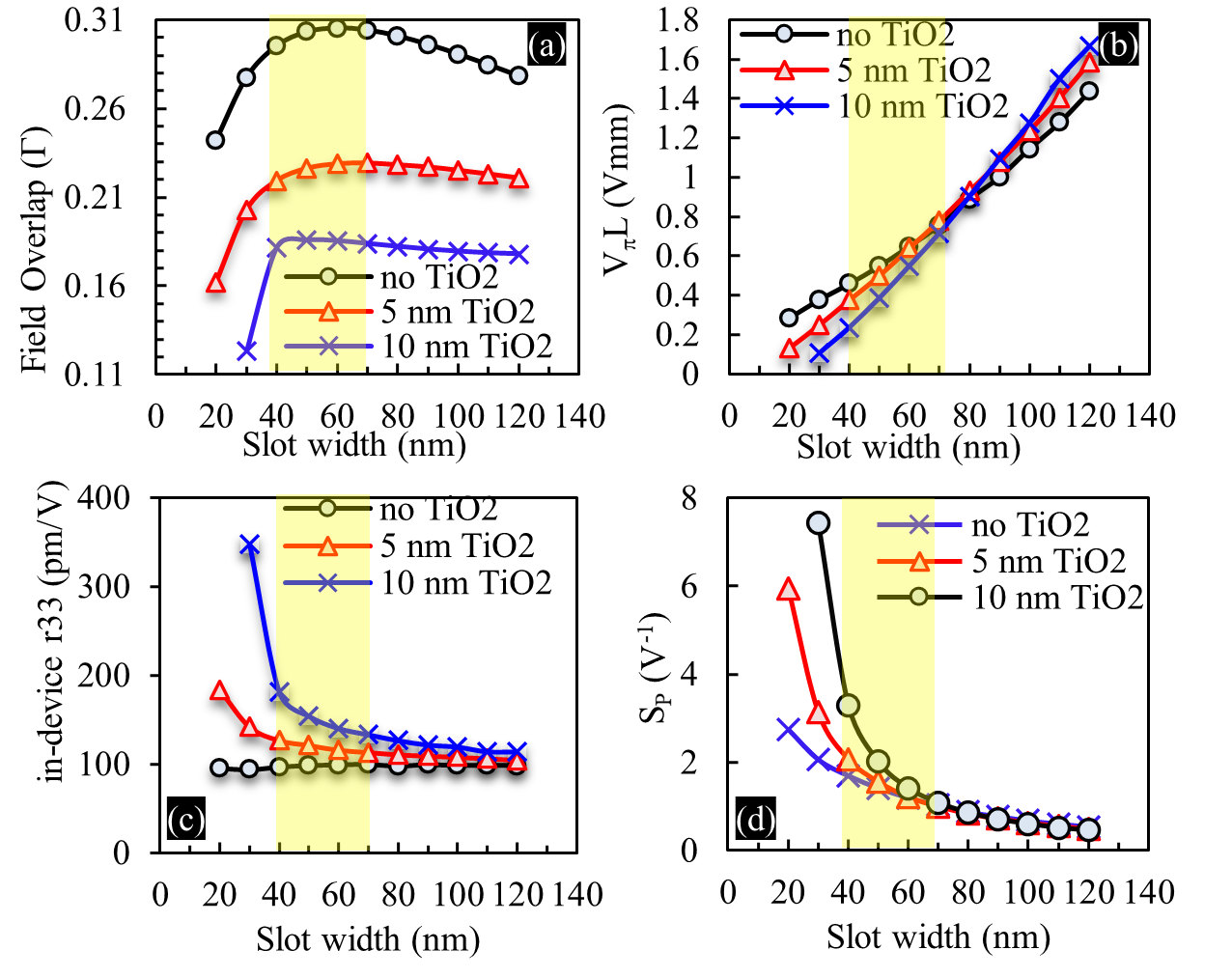}
\caption{. The (a) field overlap integral ($\Gamma$), (b) modulation efficiency ($V_\pi L$), (c) $r_{\text{33,in-device}}$, and (d) modulation sensitivity ($S_p$), all versus the slot width. Better performance is achieved by using a smaller slot, our TiO$_2$ surface treatment, or a combination of both.}
\label{FieldOverlap}
\end{figure}

Figure \ref{Transmission} displays the shift in the transmission of the MZM as a function of the applied voltage. A shift of 120 pm/V with a free-spectral range (FSR) of $\sim$ 1.15 nm is observed. The large ER > 20 dB is due to our use of directional couplers, instead of the more conventional Y-junctions or MMI couplers. As shown in Figs. \ref{ModeVsThickness}(a) and \ref{ModeVsThickness}(b), $\partial n / \partial V$ (and hence $V_\pi L$) changes in different directions, depending on the TiO$_2$ thickness. Indeed, the TiO$_2$ protection layer results in larger possible polling fields with lower currents, and hence better chromophore orientation, i.e., enhanced polling efficiency. To obtain the range of slot widths in which the performance is boosted by adding TiO$_2$, we study $r_{33,\text{in-device}}$, shown in Fig. \ref{FieldOverlap}. Rewriting Eq. (1), this factor is calculated by

\begin{equation}
r_{\text{33,in-device}} = \frac{\lambda W_{eff}}{n_e^3\Gamma V_{\pi} L}
\end{equation}

The benefits of using a 40 nm slot is evident from Fig. 13 in which low $V_\pi L$ values, high $r_{\text{33,in-device}}$ values, and large field overlap integrals are predicted. This region is indicated by the yellow band. However, for $r_{\text{33,in-device}}$ to be increased, the enhancements that the TiO$_2$ brings are required. Figure \ref{FieldOverlap}(d) illustrates the most fundamental figure of merit, i.e., the modulation sensitivity, $S_p$, while clearly showing the advantage of using the TiO$_2$ layer with thin slots. In these estimations, we used the maximum value of $r_{33} \sim$ 100 pm/v for TiO$_2$-added devices (with close-to-perfect EOP infiltration), and a correction factor of $\sim$ 0.77 was applied to the case of devices with no TiO$_2$ layer based on our previous work \cite{taghavi2018enhanced} ($V_\pi L$ = 3.125 V.mm without TiO$_2$ compared to 2.39 V.mm with 5 nm TiO$_2$, both with 40 nm slots). The advantage of the 40 nm slot along with our surface treatment is clear from Fig. \ref{FieldOverlap}. Since the modulation gain is directly proportional to $S_p$, smaller slots can simultaneously result in higher modulation gains and modulation efficiencies when appropriate TiO$_2$ layers are added to improve the polling efficiency. Indeed, another advantage of our oxide surface treatment is that it can result in higher modulation sensitivity for smaller (and hence more challenging, fabrication-wise) slot widths because of better polling (represented by $r_{\text{33,in-device}}$). In other words, instead of using a 40 nm slot, one can use a 50 nm one and add 5 nm TiO$_2$ to obtain the same level of field enhancement.\newline

We should also point out that, in this work, we only focused on enhanced polling and infiltration, so lightly doped Si has been used. Applying properly doped Si for the pedestal and waveguide ridges, one may obtain high-speed operation. As investigated in Ref. \cite{szep2011poling}, surface passivation with TiO$_2$ might result in enhanced conductivity of the pedestals, hence improving the polling efficiency and speed, indicating that we might get two benefits from applying an oxide interface. Even though we have focused on polling efficiency, our electrostatic and electric simulations (using COMSOL Multiphysics) predict a bandwidth of 39 GHz along with a significantly low modulation efficiency of $V_\pi L \approx$  0.35 V.mm when carefully doped Si is utilized. Also, the lower drive voltages obtained in this paper can result in reduced transmittance noise figures. Finally, further studies are required to determine the effectiveness of alternate passivation materials, including hafnia (HfO$_2$).

\section{Conclusion}
An MZM implemented on the SOH platform was introduced in which we utilized the well-known advantages of slot waveguides and the large electro-optic effects of polymers. To improve the modulation efficiency, slot widths as small as 40 nm were fabricated for the first time. To overcome the challenge posed by the high polling current, which adversely affects the modulation efficiency, we developed a series of fabrication steps to passivate the surface and to improve EOP infiltration based on using atomically-thin layers of TiO$_2$. Both FDTD and electrostatic simulations were employed to optimize the geometry of the surface-treated device. For narrow slots featuring TiO$_2$ passivation layers, we demonstrated a factor > 3 reduction in the polling current as well as a larger polling field, a uniformly-distributed, highly-confined, in-slot modal (optical) field. As a result, we have achieved a larger $r_{\text{33,in-device}}$ compared to the case with no surface passivation. Also, almost perfect infiltration, with a very thin (< 450 nm) final EOP layer inside the ultra-thin slot was obtained using our modified polymer spin-coating technique. This study shows that an increase in the polling/RF field inside of the 40 nm (100 nm) slots of 30\ (10\%) is easily obtained with a buffer layer of only 5 nm of TiO$_2$. In other words, without the oxide interface layer, the required polling voltage must be higher to obtain the minimum required field of 100 V/$\mu$m for the SEO125B polymer. As discussed in Ref. \cite{schulz2015mechanism}, the maximum field that can be withstood by the EOP can be greatly enhanced beyond the nominal value of 100 V/$\mu$m, thus further increasing the modulation efficiency. Also, we have studied the various figures of merit, among which is $S_p$, and we found that $S_p$ is improved from 1.7 V$^{-1}$ to 2.07 V$^{-1}$ for the 5 nm TiO$_2$ layer and from 1.7 V$^{-1}$ to 3.3 V$^{-1}$ for the 10 nm TiO$_2$ layer inside our 40 nm slot-waveguide modulators. To improve compactness, we have also developed fabrication steps to create air-bridge interconnects to carry electrical signals, instead of using conventional vias. Further studies are required to determine the effectiveness of other alternative materials including HfO$_2$. Even though we have focused on the polling efficiency, our simulations show a bandwidth of 39 GHz and a modulation efficiency of $V_\pi L \approx$ 0.35 V.mm should be obtainable. The lower drive voltages that are made possible using our approaches toward designing, and our methods for fabricating, ultra-narrow EOP-infiltrated slot waveguides will also lead to reduced transmittance noise figures for photonic integrated circuits.

\section{Author Contributions}
The main idea of the demonstrated slot modulators along with all designs and fabrication processes were developed and implemented by I.T., A.A.E., and A.A at Georgi Tech. R.D., H.M., and T.F. participated in the fabrication process implemented by I.T.,  S. S., L. C., and N. A. F. J. and A.T. participated in the simulations. All authors contributed to the manuscript All authors have given approval to the final version of the manuscript. 

\section{Acknowledgments}
The work was supported by the Defense Advanced Research Project Agency (DARPA) through the DARPA MOABB project. It was performed in part at the Georgia Tech IEN, a member of the National Nanotechnology Coordinated Infrastructure (NNCI), which is supported by the National Science Foundation (ECCS-1542174). Access to tools were facilitated by CMC Microsystems.



\bibliography{references}

\begin{thebibliography}{10}
\newcommand{\enquote}[1]{``#1''}

\bibitem{zhang2016high}
X.~Zhang, C.-J. Chung, A.~Hosseini, H.~Subbaraman, J.~Luo, A.~K. Jen, R.~L.
  Nelson, C.~Y. Lee, and R.~T. Chen, \enquote{High performance optical
  modulator based on electro-optic polymer filled silicon slot photonic crystal
  waveguide,} {\protect\JournalTitle{Journal of Lightwave Technology}}
  \textbf{34}, 2941--2951 (2016).

\bibitem{zhang2013wide}
X.~Zhang, A.~Hosseini, S.~Chakravarty, J.~Luo, A.~K.-Y. Jen, and R.~T. Chen,
  \enquote{Wide optical spectrum range, subvolt, compact modulator based on an
  electro-optic polymer refilled silicon slot photonic crystal waveguide,}
  {\protect\JournalTitle{Optics letters}} \textbf{38}, 4931--4934 (2013).

\bibitem{enami2016analysis}
Y.~Enami, H.~Nakamura, J.~Luo, and A.-Y. Jen, \enquote{Analysis of efficiently
  poled electro-optic polymer/tio2 vertical slot waveguide modulators,}
  {\protect\JournalTitle{Optics Communications}} \textbf{362}, 77--80 (2016).

\bibitem{enami2007hybrid}
Y.~Enami, C.~Derose, D.~Mathine, C.~Loychik, C.~Greenlee, R.~Norwood, T.~Kim,
  J.~Luo, Y.~Tian, A.-Y. Jen \emph{et~al.}, \enquote{Hybrid polymer/sol--gel
  waveguide modulators with exceptionally large electro--optic coefficients,}
  {\protect\JournalTitle{Nature Photonics}} \textbf{1}, 180--185 (2007).

\bibitem{koos2007nanophotonic}
C.~Koos, \emph{Nanophotonic devices for linear and nonlinear optical signal
  processing} (2007).

\bibitem{qiu2016athermal}
F.~Qiu, A.~M. Spring, H.~Miura, D.~Maeda, M.-a. Ozawa, K.~Odoi, and
  S.~Yokoyama, \enquote{Athermal hybrid silicon/polymer ring resonator
  electro-optic modulator,} {\protect\JournalTitle{ACS Photonics}} \textbf{3},
  780--783 (2016).

\bibitem{sun2011design}
X.~Sun, L.~Zhou, X.~Li, Z.~Hong, and J.~Chen, \enquote{Design and analysis of a
  phase modulator based on a metal--polymer--silicon hybrid plasmonic
  waveguide,} {\protect\JournalTitle{Applied Optics}} \textbf{50}, 3428--3434
  (2011).

\bibitem{anderson2006high}
P.~A. Anderson, B.~S. Schmidt, and M.~Lipson, \enquote{High confinement in
  silicon slot waveguides with sharp bends,} {\protect\JournalTitle{Optics
  Express}} \textbf{14}, 9197--9202 (2006).

\bibitem{chen2006subwavelength}
L.~Chen, J.~Shakya, and M.~Lipson, \enquote{Subwavelength confinement in an
  integrated metal slot waveguide on silicon,} {\protect\JournalTitle{Optics
  letters}} \textbf{31}, 2133--2135 (2006).

\bibitem{qiu2016efficiently}
F.~Qiu and S.~Yokoyama, \enquote{Efficiently poled electro-optic polymer
  modulators,} {\protect\JournalTitle{Optics express}} \textbf{24},
  19020--19025 (2016).

\bibitem{chen1997demonstration}
D.~Chen, H.~R. Fetterman, A.~Chen, W.~H. Steier, L.~R. Dalton, W.~Wang, and
  Y.~Shi, \enquote{Demonstration of 110 ghz electro-optic polymer modulators,}
  {\protect\JournalTitle{Applied Physics Letters}} \textbf{70}, 3335--3337
  (1997).

\bibitem{alloatti2014100}
L.~Alloatti, R.~Palmer, S.~Diebold, K.~P. Pahl, B.~Chen, R.~Dinu, M.~Fournier,
  J.-M. Fedeli, T.~Zwick, W.~Freude \emph{et~al.}, \enquote{100 ghz
  silicon--organic hybrid modulator,} {\protect\JournalTitle{Light: Science \&
  Applications}} \textbf{3}, e173--e173 (2014).

\bibitem{palmer2014high}
R.~Palmer, S.~Koeber, D.~L. Elder, M.~Woessner, W.~Heni, D.~Korn, M.~Lauermann,
  W.~Bogaerts, L.~Dalton, W.~Freude \emph{et~al.}, \enquote{High-speed, low
  drive-voltage silicon-organic hybrid modulator based on a binary-chromophore
  electro-optic material,} {\protect\JournalTitle{Journal of Lightwave
  Technology}} \textbf{32}, 2726--2734 (2014).

\bibitem{zhang2013polymer}
X.~Zhang, A.~Hosseini, X.~Lin, H.~Subbaraman, and R.~T. Chen,
  \enquote{Polymer-based hybrid-integrated photonic devices for silicon on-chip
  modulation and board-level optical interconnects,}
  {\protect\JournalTitle{IEEE Journal of Selected Topics in Quantum
  Electronics}} \textbf{19}, 196--210 (2013).

\bibitem{witzens2010design}
J.~Witzens, T.~Baehr-Jones, and M.~Hochberg, \enquote{Design of transmission
  line driven slot waveguide mach-zehnder interferometers and application to
  analog optical links,} {\protect\JournalTitle{Optics express}} \textbf{18},
  16902--16928 (2010).

\bibitem{hochberg2007towards}
M.~Hochberg, T.~Baehr-Jones, G.~Wang, J.~Huang, P.~Sullivan, L.~Dalton, and
  A.~Scherer, \enquote{Towards a millivolt optical modulator with nano-slot
  waveguides,} {\protect\JournalTitle{Optics Express}} \textbf{15}, 8401--8410
  (2007).

\bibitem{taylor2005radiation}
E.~W. Taylor, J.~E. Nichter, F.~D. Nash, F.~Haas, A.~A. Szep, R.~J. Michalak,
  B.~M. Flusche, P.~R. Cook, T.~A. McEwen, B.~F. McKeon \emph{et~al.},
  \enquote{Radiation resistance of electro-optic polymer-based modulators,}
  {\protect\JournalTitle{Applied Physics Letters}} \textbf{86}, 201122 (2005).

\bibitem{chen2011achieving}
A.~Chen, H.~Sun, A.~Szep, S.~Shi, D.~Prather, Z.~Lin, R.~S. Kim, and
  D.~Abeysinghe, \enquote{Achieving higher modulation efficiency in
  electrooptic polymer modulator with slotted silicon waveguide,}
  {\protect\JournalTitle{Journal of lightwave technology}} \textbf{29},
  3310--3318 (2011).

\bibitem{wang2011effective}
X.~Wang, C.-Y. Lin, S.~Chakravarty, J.~Luo, A.~K.-Y. Jen, and R.~T. Chen,
  \enquote{Effective in-device r 33 of 735 pm/v on electro-optic polymer
  infiltrated silicon photonic crystal slot waveguides,}
  {\protect\JournalTitle{Optics letters}} \textbf{36}, 882--884 (2011).

\bibitem{szep2011poling}
A.~Szep, A.~Chen, S.~Shi, Z.~Lin, and D.~Abeysinghe, \enquote{Poling study of
  electro-optic polymers in silicon slot waveguides,} in \emph{RF and
  Millimeter-Wave Photonics,}  vol. 7936 (International Society for Optics and
  Photonics, 2011), p. 79360C.

\bibitem{huang2012efficient}
S.~Huang, J.~Luo, H.-L. Yip, A.~Ayazi, X.-H. Zhou, M.~Gould, A.~Chen,
  T.~Baehr-Jones, M.~Hochberg, and A.~K.-Y. Jen, \enquote{Efficient poling of
  electro-optic polymers in thin films and silicon slot waveguides by
  detachable pyroelectric crystals,} {\protect\JournalTitle{Advanced
  Materials}} \textbf{24}, OP42--OP47 (2012).

\bibitem{schulz2015mechanism}
K.~M. Schulz, S.~Prorok, D.~Jalas, S.~R. Marder, J.~Luo, A.~K.-Y. Jen,
  R.~Zierold, K.~Nielsch, and M.~Eich, \enquote{Mechanism that governs the
  electro-optic response of second-order nonlinear polymers on silicon
  substrates,} {\protect\JournalTitle{Optical Materials Express}} \textbf{5},
  1653--1660 (2015).

\bibitem{baehr2008nonlinear}
T.~Baehr-Jones, B.~Penkov, J.~Huang, P.~Sullivan, J.~Davies, J.~Takayesu,
  J.~Luo, T.-D. Kim, L.~Dalton, A.~Jen \emph{et~al.}, \enquote{Nonlinear
  polymer-clad silicon slot waveguide modulator with a half wave voltage of
  0.25 v,} {\protect\JournalTitle{Applied Physics Letters}} \textbf{92}, 147
  (2008).

\bibitem{tang1997enhanced}
H.~Tang, J.~M. Taboada, G.~Cao, L.~Li, and R.~T. Chen, \enquote{Enhanced
  electro-optic coefficient of nonlinear optical polymer using liquid contact
  poling,} {\protect\JournalTitle{Applied physics letters}} \textbf{70},
  538--540 (1997).

\bibitem{enami2014enhanced}
Y.~Enami, Y.~Jouane, J.~Luo, and A.~K. Jen, \enquote{Enhanced conductivity of
  sol-gel silica cladding for efficient poling in electro-optic polymer/tio 2
  vertical slot waveguide modulators,} {\protect\JournalTitle{Optics express}}
  \textbf{22}, 30191--30199 (2014).

\bibitem{li2015poling}
M.~Li, S.~Huang, X.-H. Zhou, Y.~Zang, J.~Wu, Z.~Cui, J.~Luo, and A.~K.-Y. Jen,
  \enquote{Poling efficiency enhancement of tethered binary nonlinear optical
  chromophores for achieving an ultrahigh n 3 r 33 figure-of-merit of 2601 pm
  v- 1,} {\protect\JournalTitle{Journal of Materials Chemistry C}} \textbf{3},
  6737--6744 (2015).

\bibitem{jouane2014unprecedented}
Y.~Jouane, Y.~Chang, D.~Zhang, J.~Luo, A.~K. Jen, and Y.~Enami,
  \enquote{Unprecedented highest electro-optic coefficient of 226 pm/v for
  electro-optic polymer/tio2 multilayer slot waveguide modulators,}
  {\protect\JournalTitle{Optics express}} \textbf{22}, 27725--27732 (2014).

\bibitem{sprave1996high}
M.~Sprave, R.~Blum, and M.~Eich, \enquote{High electric field conduction
  mechanisms in electrode poling of electro-optic polymers,}
  {\protect\JournalTitle{Applied physics letters}} \textbf{69}, 2962--2964
  (1996).

\bibitem{blum1998high}
R.~Blum, M.~Sprave, J.~Sablotny, and M.~Eich, \enquote{High-electric-field
  poling of nonlinear optical polymers,} {\protect\JournalTitle{JOSA B}}
  \textbf{15}, 318--328 (1998).

\bibitem{lou2013design}
F.~Lou, D.~Dai, L.~Thylen, and L.~Wosinski, \enquote{Design and analysis of
  ultra-compact eo polymer modulators based on hybrid plasmonic microring
  resonators,} {\protect\JournalTitle{Optics express}} \textbf{21},
  20041--20051 (2013).

\bibitem{korn2014electro}
D.~Korn, M.~Jazbinsek, R.~Palmer, M.~Baier, L.~Alloatti, H.~Yu, W.~Bogaerts,
  G.~Lepage, P.~Verheyen, P.~Absil \emph{et~al.}, \enquote{Electro-optic
  organic crystal silicon high-speed modulator,} {\protect\JournalTitle{IEEE
  Photonics Journal}} \textbf{6}, 1--9 (2014).

\bibitem{lee2011recent}
Y.~Lee, S.~Takahashi, H.~Mahalingam, W.~Steier, G.~Betts, and J.~Chen,
  \enquote{Recent progress toward a nanoslot modulator: polymer poling
  experiments,} in \emph{RF and Millimeter-Wave Photonics,}  vol. 7936
  (International Society for Optics and Photonics, 2011), p. 793608.

\bibitem{taghavi2018enhanced}
I.~Taghavi, R.~Dehghannasiri, T.~Fan, H.~Moradinejad, H.~Taghinejad, A.~H.
  Hosseinnia, A.~A. Eftekhar, and A.~Adibi, \enquote{Enhanced polling and
  infiltration of highly-linear mach-zehnder modulators on si/sin-organic
  hybrid platform,} in \emph{CLEO: Science and Innovations,}  (Optical Society
  of America, 2018), pp. SM1I--1.

\end{thebibliography}

\end{document}